\documentclass[12pt,prd,onecolumn,showpacs,amsmath,amssymb,aps,floats,floatfix,nofootinbib]{revtex4-1}

\usepackage[colorlinks=true,urlcolor=blue,anchorcolor=blue,citecolor=blue,filecolor=blue,linkcolor=blue,menucolor=blue,linktocpage=true]{hyperref}

\usepackage[inline]{enumitem}
\usepackage[multidot]{grffile} 
\usepackage{dcolumn}
\usepackage{bm}
\usepackage{amsmath}
\usepackage{amsfonts}
\usepackage{amssymb}
\usepackage{color}
\usepackage{latexsym}
\usepackage{slashed} 
\usepackage{pstricks}
\usepackage{indentfirst}
\usepackage{mathrsfs}
\usepackage{multirow}
\usepackage{epsfig,psfrag}
\usepackage{subfigure}
\usepackage{mathtools}
\usepackage{setspace} 
\usepackage[utf8]{inputenc} 
\usepackage[scientific-notation=true]{siunitx} 
\usepackage{verbatim}

\graphicspath{{fig/}}

\begin{document}

\title{Geminga contribution to the cosmic-ray positron excess according to the gamma-ray observations}

\author{Guang-Yao Zhou$^1$}

\author{Zhao-Huan Yu$^1$}
\email{yuzhaoh5@mail.sysu.edu.cn}

\author{Qiang Yuan$^{2,3}$}
\email{yuanq@pmo.ac.cn}

\author{Hong-Hao Zhang$^1$}
\email{zhh98@mail.sysu.edu.cn}

\affiliation{$^1$School of Physics, Sun Yat-Sen University, Guangzhou 510275, China}

\affiliation{$^2$Key Laboratory of Dark Matter and Space Astronomy, Purple Mountain Observatory, Chinese Academy of Sciences, Nanjing 210023, China}
\affiliation{$^3$School of Astronomy and Space Science, University of Science and Technology of China, Hefei 230026, China}

\begin{abstract}
We attempt to interpret the cosmic-ray positron excess by injection from the nearby pulsar Geminga, assuming a two-zone diffusion scenario and an injection spectrum with a low energy cutoff. Since the high energy positrons and electrons from Geminga can induce $\gamma$ rays via inverse Compton scattering, we take into account the extended $\gamma$-ray observations around Geminga from HAWC for $\sim 10$~TeV and from Fermi-LAT for $\mathcal{O}(10)$~GeV. According to the extended $\gamma$-ray observation claimed by an analysis of Fermi-LAT data, we find that Geminga could explain the positron excess for a $30\%$ energy conversion efficiency into positrons and electrons. However, based on the constraint on the extended $\gamma$ rays given by another Fermi-LAT analysis, positrons from Geminga would be insufficient to account for the positron excess. A further robust analysis of Fermi-LAT data for the extended $\gamma$ rays would be crucial to determine whether Geminga can explain the positron excess or not.
\end{abstract}

\maketitle
\tableofcontents

\section{Introduction}
\label{sec:intro}

Since 2008, the measurements of cosmic-ray (CR) positrons by PAMELA~\cite{PAMELA:2008gwm}, Fermi-LAT~\cite{Fermi-LAT:2011baq}, and AMS-02~\cite{AMS:2013fma, AMS:2019iwo} have shown an unexpected excess at energies $\gtrsim 10~\si{GeV}$.
Possible interpretations for this excess include annihilating/decaying dark matter~\cite{Bergstrom:2008gr, Cholis:2008hb, Yin:2008bs, Yuan:2013eja} and astrophysical sources like nearby pulsars within kpc~\cite{Hooper:2008kg, Yuksel:2008rf}.
In particular, the middle-aged pulsar Geminga with a distance of $\sim 250~\si{pc}$ is widely assumed to produce high energy positrons that could propagate to the Earth~\cite{Hooper:2008kg, Yuksel:2008rf, Yin:2013vaa, Feng:2015uta, Hooper:2017tkg, Cholis:2017ccs, Fang:2018qco, Profumo:2018fmz, Cholis:2018izy, Tang:2018wyr, Xi:2018mii, Johannesson:2019jlk, DiMauro:2019yvh, Fang:2019ayz, Manconi:2020ipm, Wang:2021xph, Fang:2022mdg}.

In 2017, the HAWC experiment observed $\sim 10~\si{TeV}$ $\gamma$ rays spatially extended about 2 degrees around Geminga, which would be produced by positrons and electrons of energies $\sim 100~\si{TeV}$ via inverse Compton scattering (ICS) off low energy photons~\cite{HAWC:2017kbo}.
Therefore, this observation confirms that Geminga is a source of high energy positrons and electrons.
But the surface brightness profile (SBP) measured by HAWC implies a diffusion coefficient smaller than the conventional value by at least two orders of magnitude.
The recent observation of another extended halo around the middle-aged pulsar J0621+3749 by
LHAASO further established the general conclusion of slow diffusion around pulsars
\cite{LHAASO:2021crt}. 
Such slow diffusion results in much less CR positrons arriving at the Earth, unlikely to explain the positron excess.
Nonetheless, by assuming a two-zone diffusion model with slow diffusion in a small zone around the source but normal diffusion outside the zone, positrons originated from Geminga can still sufficiently contribute to the positron excess~\cite{Fang:2018qco,Profumo:2018fmz,Bao:2021hey}.

In addition, positrons and electrons from Geminga are also expected to induce extended ICS $\gamma$ rays in the energy range of Fermi-LAT.
Based on two-zone diffusion templates, an analysis of 10-yr Fermi-LAT $\gamma$-ray data by Xi et al.~\cite{Xi:2018mii} (denoted as X19 below) did not find such extended emission and derive a stringent constraint on the $\gamma$-ray flux in the $\sim 5\text{--}100$~\si{GeV} energy range.
According to this constraint and the HAWC data,  $e^\pm$ from Geminga with a single power-law injection spectrum can only contribute a small faction to the CR positron spectrum observed by AMS-02.

On the other hand, taking into account both a larger region of interest and the proper motion of the Geminga pulsar, another analysis of Fermi-LAT data by Di Mauro et al.~\cite{DiMauro:2019yvh} (denoted as D19 hereafter) claimed a discovery of extended $\gamma$-ray emissions around Geminga in the energy range of $\sim 10$--$100$~\si{GeV}.
However, considering both the corresponding $\gamma$-ray flux and the HAWC data, the Geminga contribution to the position flux they obtained is not enough for the AMS-02 excess.

Both the X19 and D19 analyses assumed a single power-law Geminga $e^\pm$ injection spectrum with a high energy cutoff.
The inconsistency with the AMS-02 data may indicate that there are less low energy positrons and electrons producing GeV $\gamma$ rays.
Therefore, we will attempt to modify the injection spectrum by adding a low energy cutoff, in order to simultaneously explain the HAWC, Fermi-LAT, and AMS-02 data.
The results of the $\gamma$-ray flux from the X19 and D19 analyses will be considered separately.

This paper is organized as follows.
In Section~\ref{sec:Geminga}, we describe the propagation of positrons and electrons produced by Geminga and the $\gamma$-ray flux induced by ICS.
In Section~\ref{sec:D19}, we simultaneously interpret the HAWC data, the Fermi-LAT $\gamma$-ray observation given by D19, and the AMS-02 positron spectrum assuming an $e^\pm$ injection spectrum with a low energy cutoff.
In Section~\ref{sec:X19}, we use the Fermi-LAT $\gamma$-ray constraint given by X19 to explore how much contribution Geminga can supply to the AMS-02 positron excess.
Section~\ref{sec:sum} gives the summary and discussion.

\section{Positrons and electrons from Geminga}
\label{sec:Geminga}

The Geminga pulsar is a $\gamma$-ray source discovered by SAS-2~\cite{Fichtel:1975}.
Its age is about $342~\si{kyr}$~\cite{Manchester:2004bp}, and the distance for the Earth is $250^{+120}_{-62}~\si{pc}$~\cite{Faherty:2007}.
Geminga is expected to emit lots of positrons and electrons, which diffuse away from Geminga and lose energies by upscatter low energy photons in the cosmic microwave background (CMB) and interstellar radiation backgrounds through ICS processes.

The propagation of CR $e^\pm$ is described by the diffusion-cooling
equation
\begin{equation}
\frac{\partial N}{\partial t} - \nabla\cdot(D\nabla N) - 
\frac{\partial}{\partial
	E}(bN) = Q \,,
\label{eq:prop}
\end{equation}
where $N$ is the $e^\pm$ differential density, $E$ is the $e^\pm$ energy, $D$ is the diffusion coefficient, and $Q$ is the source term.
The energy loss rate $b$ includes both contributions from synchrotron radiation and ICS.
The synchrotron energy loss rate in a magnetic field $B$ is given by~\cite{Crusius:1988}
\begin{equation}
b_\mathrm{syn} = \frac{4 \sigma_\mathrm{T} \gamma_e^2 U_B}{3 m_e c},
\end{equation}
where $\sigma_\mathrm{T}$ is the Thomson cross section, $\gamma_e = E/(m_e c^2)$ is the $e^\pm$ Lorentz factor, and $U_B = B^2/(8\pi)$ is the energy density of the magnetic field.
The energy loss rate due to ICS is estimated following Ref.~\cite{Fang:2020dmi}.
We convert the propagation equation to a difference equation, which is solved using the numerical method described in Ref.~\cite{Fang:2018qco}.

We assume a spherically symmetrical two-zone diffusion scenario with the diffusion coefficient given by
\begin{equation}
D(E, r)=\left\{
\begin{aligned}
D_1(E), & & r< r_\star, \\
D_2(E), & & r\geq r_\star.
\end{aligned}
\right.
\label{eq:diff}
\end{equation}
Here $r$ is the distance from Geminga, and $r_\star$ denotes the boundary of the two diffusion zones.
Both $D_1(E)$ and $D_2(E)$ are assumed to have a form of $D_{100} (E/100~\si{TeV})^\delta$, where $D_{100}$ is the diffusion coefficient at $E = 100~\si{TeV}$, and $\delta = 0.33$ is adopted for a Kolmogorov-type diffusion~\cite{Kolmogorov:1941}.

The morphological SBP study of the extended TeV $\gamma$-ray emissions around Geminga by HAWC gives a diffusion coefficient $D_{100} = (3.2^{+1.4}_{-1.0})\times10^{27}~\si{cm^{2}~s^{-1}}$ for $100~\si{TeV}$ $e^\pm$ around Geminga, while a similar study of another nearby pulsar Monogen (PSR B0656+14) leads to $D_{100} = (15^{+49}_{-9})\times10^{27}~\si{cm^{2}~s^{-1}}$~\cite{HAWC:2017kbo}.
The joint fit of both results in $D_{100} = (4.5\pm 1.2)\times10^{27}~\si{cm^{2}~s^{-1}}$.
Thus, $D_{100}$ for the inner zone with $r< r_\star$ is at the order of $10^{27}~\si{cm^{2}~s^{-1}}$.
For the outer zone with $r \geq r_\star$, positrons and electrons propagate through the ordinary interstellar medium (ISM), and we take the GALPROP~\cite{Moskalenko:1997gh} default value $D_{100} = \num{1.7e30}~\si{cm^{2}~s^{-1}}$, 
which is consistent with the measurements of CR secondary-to-primary ratios, particularly the $\mathrm{B}/\mathrm{C}$ ratio.

The source term for high energy $e^\pm$ injected by Geminga is assumed as
\begin{equation}
Q(t,E,r) = q(t,E)\delta(r)\,,
\label{eq:source}
\end{equation}
where
\begin{equation}
q(t,E)=q_0\left(1+\frac{t}{\tau}\right)^{-2}E^{-\gamma} \exp\left(-\frac{E}{E_\mathrm{hc}}\right) \exp\left(-\frac{E_\mathrm{lc}}{E}\right).
\label{eq:time_profile}
\end{equation}
$\tau$ is the characteristic initial spin-down time scale of the Geminga pulsar, taken to be $12~\si{kyr}$ following Ref.~\cite{HAWC:2017kbo}.
$\gamma$ is the injection spectral index for $e^\pm$.
$E_\mathrm{hc}$ and $E_\mathrm{lc}$ are the high and low energy cutoffs, respectively.
$q_0$ is a constant determined by the normalization relation
\begin{equation}
\int_{E_{\rm min}}^{E_{\rm max}}q(t_\mathrm{s},E)E\mathrm{d}E=\eta\dot{E}_s\,,
\label{eq:norm}
\end{equation}
with the Geminga age $t_\mathrm{s}=342~\si{kyr}$ and the spin-down luminosity $\dot{E}_\mathrm{s}=\num{3.2e34}~\si{erg~s^{-1}}$~\cite{Manchester:2004bp}.
Here $\eta$ is the conversion efficiency for the spin-down energy converted to $e^\pm$ energies.
We will not consider the difference between the positrons and electrons when calculating the $\gamma$-ray flux, and the positron flux $\Phi_{e^+}$ is just a half of the total $e^\pm$ flux.

The photon emissivity due to $e^\pm$ ICS based on the Klein-Nishina cross section is given by~\cite{Fang:2007sc}
\begin{equation}
Q_\mathrm{ICS} (t, E_\gamma, r) = 4\pi \sum_j \int^{\infty}_{0}\mathrm{d}\epsilon
\, n_j(\epsilon)\int^{E_{\mathrm{max}}}_{E_{\mathrm{min}}}\mathrm{d}E \,
J(t, E, r)F(\epsilon, E_{\gamma}, E),
\end{equation}
$n_j(\epsilon)$ is the number density of a background photon component $j$ with energy $\epsilon$, temperature $T_j$, and energy density $U_j$, expressed as
\begin{equation}
n_j(\epsilon)=\frac {15U_j}{(\pi k
	T_j)^4}\frac{\epsilon^2}{\exp(\epsilon/kT_j)-1},
\end{equation}
where $k$ is the Boltzmann constant.
The $e^\pm$ energy threshold for upscattering a target photon with energy $\epsilon$ to a photon with energy $E_\gamma$ is
\begin{equation}
E_\mathrm{min} = \frac{1}{2} \left( E_\gamma + \sqrt{E_\gamma^2 + \frac{E_\gamma m_e^2 c^4 }{\epsilon}} \right).
\end{equation}
$J(t, E, r) = v_e N(t, E, r)/(4\pi)$ is the $e^\pm$ intensity, with $v_e$ denoting the $e^\pm$ speed, which approaches the light speed $c$ for high energy $e^\pm$.
The function $F$ is given by
\begin{equation}
F(\epsilon, E_{\gamma}, E) = \frac{3\sigma_\mathrm{T}}{4\gamma_e^2 \epsilon}
\left[ 2q\ln q + (1+2q)(1-q) + \frac{\Gamma^2 q^2(1-q)}{2(1+\Gamma q)}
\right],
\end{equation}
with
\begin{equation}
\Gamma = \frac{4\epsilon\gamma_e}{m_e c^2},\quad
q = \frac{E_\gamma}{\Gamma (E_e - E_\gamma)}.
\end{equation}

Following Ref.~\cite{HAWC:2017kbo}, we consider three background photon components, including the CMB, the IR background, and the optical background, for the ICS processes.
The temperatures and energy densities are presented in Table~\ref{tab:ISRF}.
Integrating $Q_\mathrm{ICS} (t_s, E_\gamma, r)$ along the light of sight~\cite{Liu:2019sfl}, we obtain the $\gamma$-ray flux for specific energy $E_\gamma$ and angular separation $\theta$,
\begin{equation}
I(E_\gamma,\theta)=\frac{1}{4\pi}\int_{l_\mathrm{min}}^{l_\mathrm{max}}\mathrm{d}l \,Q_\mathrm{ICS}(t_s, E_\gamma, r).
\end{equation}
Then we integrate out $\theta$ to get the energy spectrum of the $\gamma$-ray flux $\Phi_\gamma$, or  integrate out $E_\gamma$ to derive the SBP as a function of $\theta$.
The angular separation $\theta$ is integrated up to $20^\circ$, which is consistent with the large regions of interest considered in the X19~\cite{Xi:2018mii} and D19~\cite{DiMauro:2019yvh} analyses of Fermi-LAT data.

\begin{table}[!t]   
\begin{center}
\setlength{\tabcolsep}{1em}
\renewcommand\arraystretch{1.3}
\caption{Temperature $T_j$ and energy density $U_j$ of three background photon components~\cite{HAWC:2017kbo}.} 
\label{tab:ISRF}
\vspace{1em}
\begin{tabular}{ccc}
\hline\hline
Component $j$ & $T_j$ (K) & $U_j$ ($\si{eV/cm^3}$)\\
\hline  
CMB& 2.7 & 0.26\\
IR & 20 & 0.3\\
Optical & 5000 & 0.3\\ 
\hline\hline
\end{tabular}  
\end{center}  
\end{table}

\section{Result according to the D19 gamma-ray observation}
\label{sec:D19}

In this section, we try to interpret the HAWC and AMS-02 data according to the Fermi-LAT $\gamma$-ray observation from the D19 analysis~\cite{DiMauro:2019yvh}.
Both the results without and with the low energy cutoff $E_\mathrm{lc}$ in the $e^\pm$ injection spectrum are calculated for comparison.

\begin{figure}[!t]
\centering
\subfigure[~$\gamma$-ray spectrum \label{fig:DM:gamma}]{\includegraphics[width=0.48\textwidth]{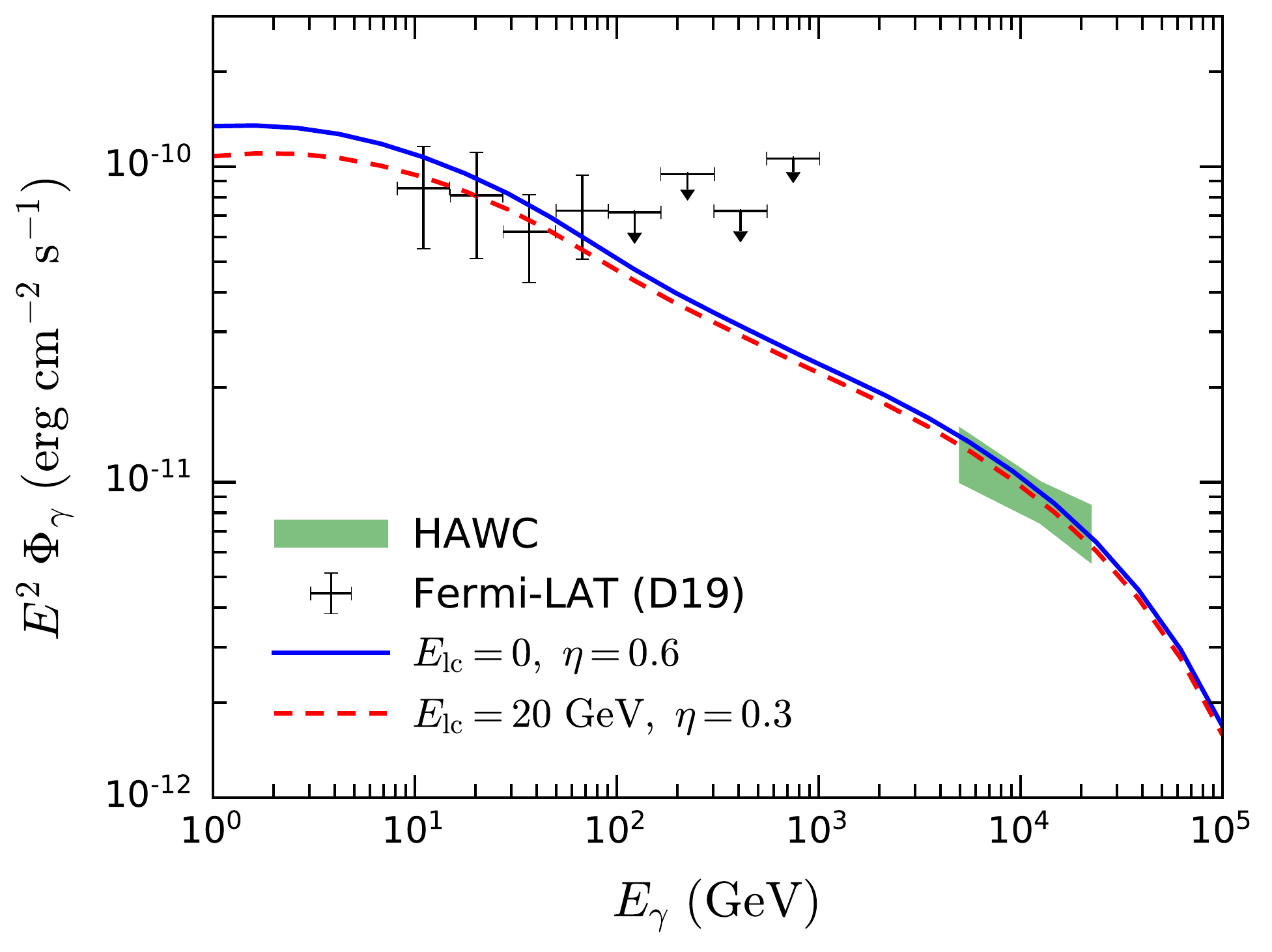}}
\subfigure[~Surface brightness profile~\label{fig:DM:SPB}]{\includegraphics[width=0.465\textwidth]{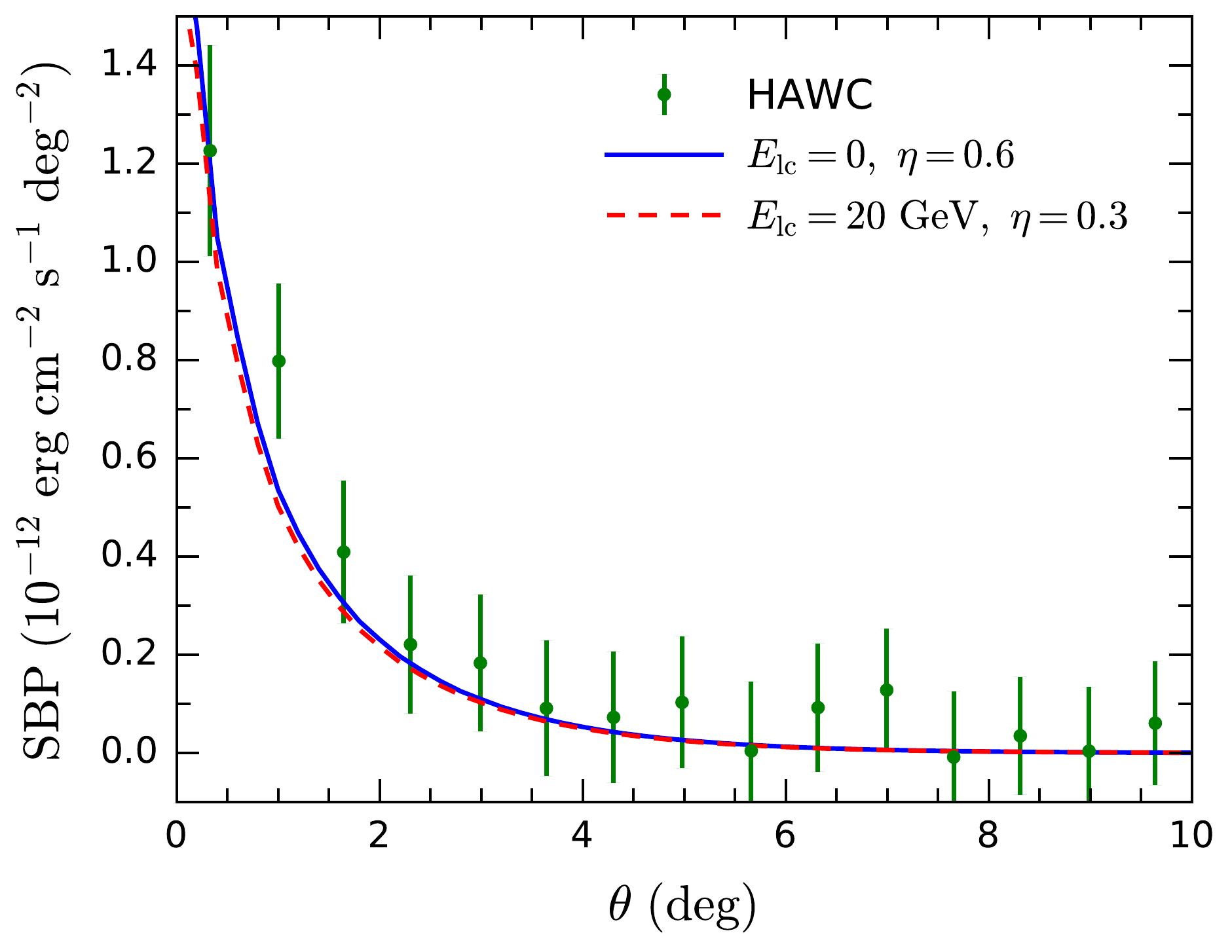}}
\subfigure[~CR positron spectrum\label{fig:DM:posi}]{\includegraphics[width=0.48\textwidth]{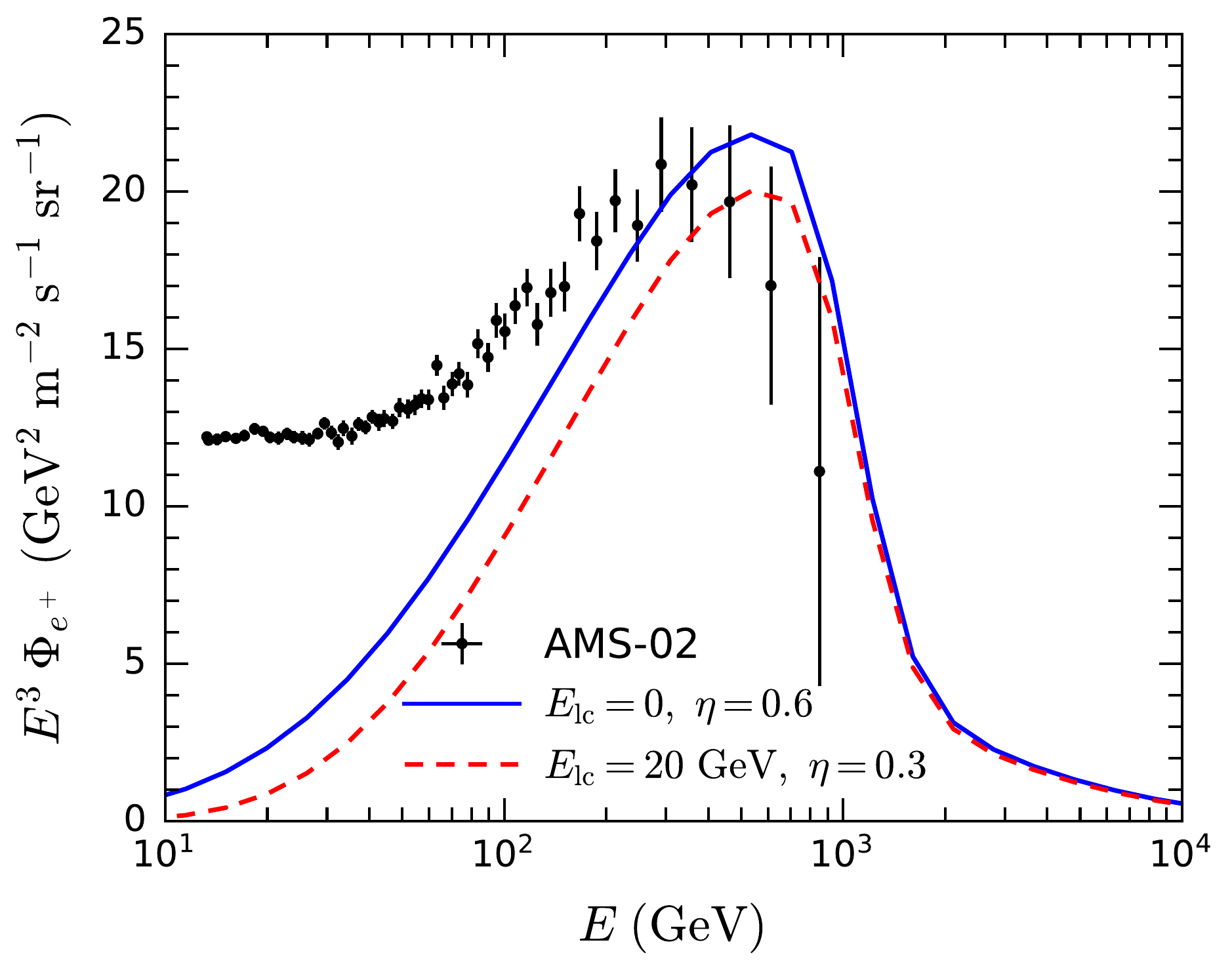}}
\caption{The $\gamma$-ray spectrum around Geminga (a), the Geminga SBP (b), and the CR positron spectrum (c) assuming $e^\pm$ injection spectra without a low energy cutoff for $\eta = 0.6$ (blue solid lines) and with $E_\mathrm{lc} = 20~\si{GeV}$ for $\eta=0.3$ (red dashed lines).
In the upper left panel, the green region denotes the $\sim 10~\si{TeV}$ spectral data measured by HAWC~\cite{HAWC:2017kbo}, and the data points and upper limits in $10~\si{GeV} \lesssim E_\gamma \lesssim \si{TeV}$ are given by the D19 analysis of Fermi-LAT data~\cite{DiMauro:2019yvh}.
The data points in the upper right panel shows the HAWC observation of the Geminga SBP~\cite{HAWC:2017kbo}.
The lower panel displays the positron spectrum measured by AMS-02~\cite{AMS:2019iwo}.}
\label{fig:DM}
\end{figure}

Firstly, we consider an $e^\pm$ injection spectrum without $E_\mathrm{lc}$, and adjusting the energy conversion efficiency $\eta$ to meet the data.
Setting the boundary radius $r_\star = 50~\si{pc}$, the diffusion coefficient at $E = 100~\si{GeV}$ in the inner diffusion zone $D_{100} = 3.5\times10^{27}~\si{cm^{2}~s^{-1}}$, the ISM magnetic field $B = 3~\si{\mu G}$, the $e^\pm$ injection spectral index $\gamma = 2.2$, and the high energy cutoff $E_\mathrm{hc} = 511~\si{TeV}$, we derive the $\gamma$-ray spectrum around Geminga, the Geminga SBP, and the CR positron spectrum at the Earth for $\eta = 0.6$, shown as the blue solid lines in Fig.~\ref{fig:DM}.

In order to compare the predictions and the observations, we show the $\sim 10~\si{TeV}$ spectral data measured by HAWC~\cite{HAWC:2017kbo} and the Fermi-LAT data points and upper limits from $\sim 10~\si{GeV}$ to $\sim \si{TeV}$ given by the D19 analysis~\cite{DiMauro:2019yvh} in Fig.~\ref{fig:DM:gamma}.
The HAWC observation of the Geminga SBP~\cite{HAWC:2017kbo} is demonstrated in Fig.~\ref{fig:DM:SPB}, while the positron spectrum measured by AMS-02~\cite{AMS:2019iwo} is displayed in Fig.~\ref{fig:DM:posi}.
For the above setup with $\eta = 0.6$, we find that the $\gamma$-ray prediction can well interpret the $\gamma$-ray spectrum and the SBP, and the predicted $e^+$ spectrum can explain the AMS-02 data at $E \gtrsim 100~\si{GeV}$.
However, a $60\%$ efficiency of the spin-down energy converted to $e^\pm$ energies looks unrealistic.

Secondly, we introduce a low energy cutoff $E_\mathrm{lc} = 20~\si{GeV}$ in the $e^\pm$ injection spectrum with other parameters unchanged, and find that the observational data can be explained for $\eta = 0.3$, as illustrated as the red dashed lines in Fig.~\ref{fig:DM}.
Such a $30\%$ conversion efficiency is much more reasonable than the previous one.
Now the predicted positron flux at $E \lesssim 100~\si{GeV}$ seems slightly lower than the blue solid line, but we can still interpret the AMS-02 data at $E \gtrsim 100~\si{GeV}$ very well.

\section{Result according to the X19 gamma-ray constraint}
\label{sec:X19}

In contrast to the D19 analysis~\cite{DiMauro:2019yvh}, the X19 analysis of the Fermi-LAT data have not found any extended $\gamma$-ray emission around Geminga, deriving a rather stringent constraint on the $\gamma$-ray flux at $\sim 5\text{--}100~\si{GeV}$~\cite{Xi:2018mii}.
In this section, we consider this constraint to see how it affects the Geminga contribution to the CR positron spectrum, assuming a low energy cutoff in the $e^\pm$ injection spectrum.
However,  we find it impossible to simultaneously explain the HAWC, Fermi-LAT, and AMS-02 data, because the X19 constraint is too strict.
Instead, we would like to know how much contribution Geminga can provide to the AMS-02 positron excess.

For this purpose, we treat $\gamma$, $E_\mathrm{hc}$, $E_\mathrm{lc}$, $\eta$, $B$, and $D_{100}$ in the inner diffusion zone as free parameters and perform a scan in the parameter space with fixed $r_\star$, utilizing the \texttt{MultiNest} algorithm~\cite{Feroz:2008xx} to improve the fitting efficiency.
The ranges for the free parameters in the scan are chosen to be
\begin{eqnarray}
&& 1.8 < \gamma < 2.2,\quad
200~\si{TeV} < E_\mathrm{hc} < 600~\si{TeV},\quad
100~\si{GeV} < E_\mathrm{lc} < 900~\si{GeV},
\nonumber\\
&& 0.1 < \eta < 0.4,\quad
3~\si{\mu G} < B < 8~\si{\mu G},\quad
10^{26}~\si{cm^2~s^{-1}} < D_{100} < 10^{27}~\si{cm^2~s^{-1}}.
\end{eqnarray}

\begin{table}[!t]
\begin{center}
\setlength{\tabcolsep}{1em}
\renewcommand\arraystretch{1.3}
\caption{Parameters in the best results for fixed $r_\star$.}
\label{tab:bestpara}
\vspace{1em}
\begin{tabular}{cccc}
\hline\hline
$r_\star$ (pc) & $50$ & $70$ & $100$ \\
\hline
$\gamma$ & $2.10$ & $1.93$ & $1.70$\\
$E_\mathrm{hc}$ (TeV) & $520$ & $537$ & $463$\\
$E_\mathrm{lc}$ (GeV) & $870$ & $302$ & $547$ \\
$\eta$ & $0.15$ & $0.21$ & $0.16$ \\	
$B$ ($\si{\mu G}$) & $5.0$ & $6.9$ & $7.2$\\	
$D_{100}$ ($10^{27}~\si{cm^2~s^{-1}}$) & $4.8$ & $7.8$ & $8.5$\\
\hline\hline	 
\end{tabular}
\end{center}
\end{table}

\begin{figure}[!t]
\centering
\subfigure[~$\gamma$-ray spectrum \label{fig:Xi:gamma}]{\includegraphics[width=0.48\textwidth]{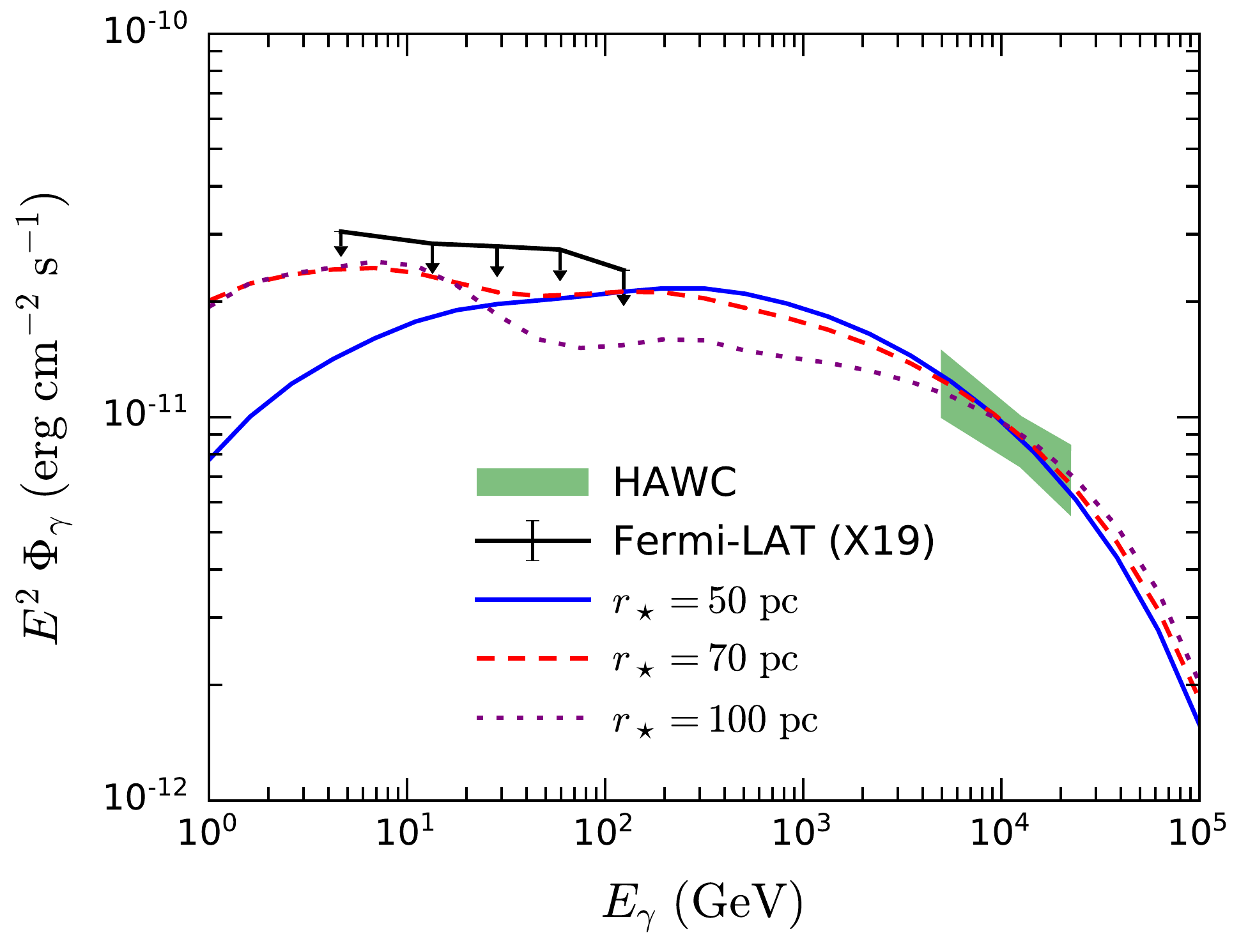}}
\subfigure[~Surface brightness profile~\label{fig:Xi:SPB}]{\includegraphics[width=0.465\textwidth]{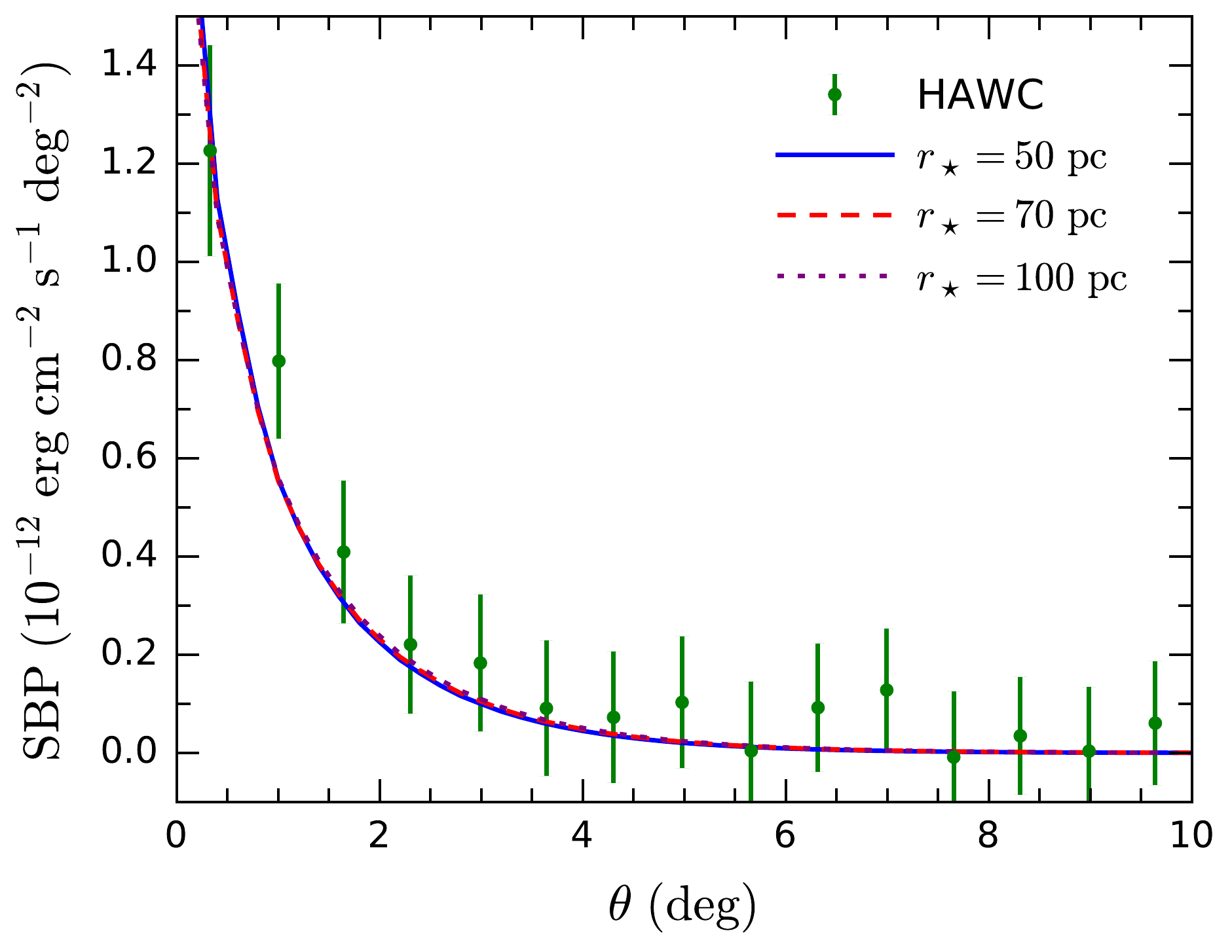}}
\subfigure[~CR positron spectrum\label{fig:Xi:posi}]{\includegraphics[width=0.48\textwidth]{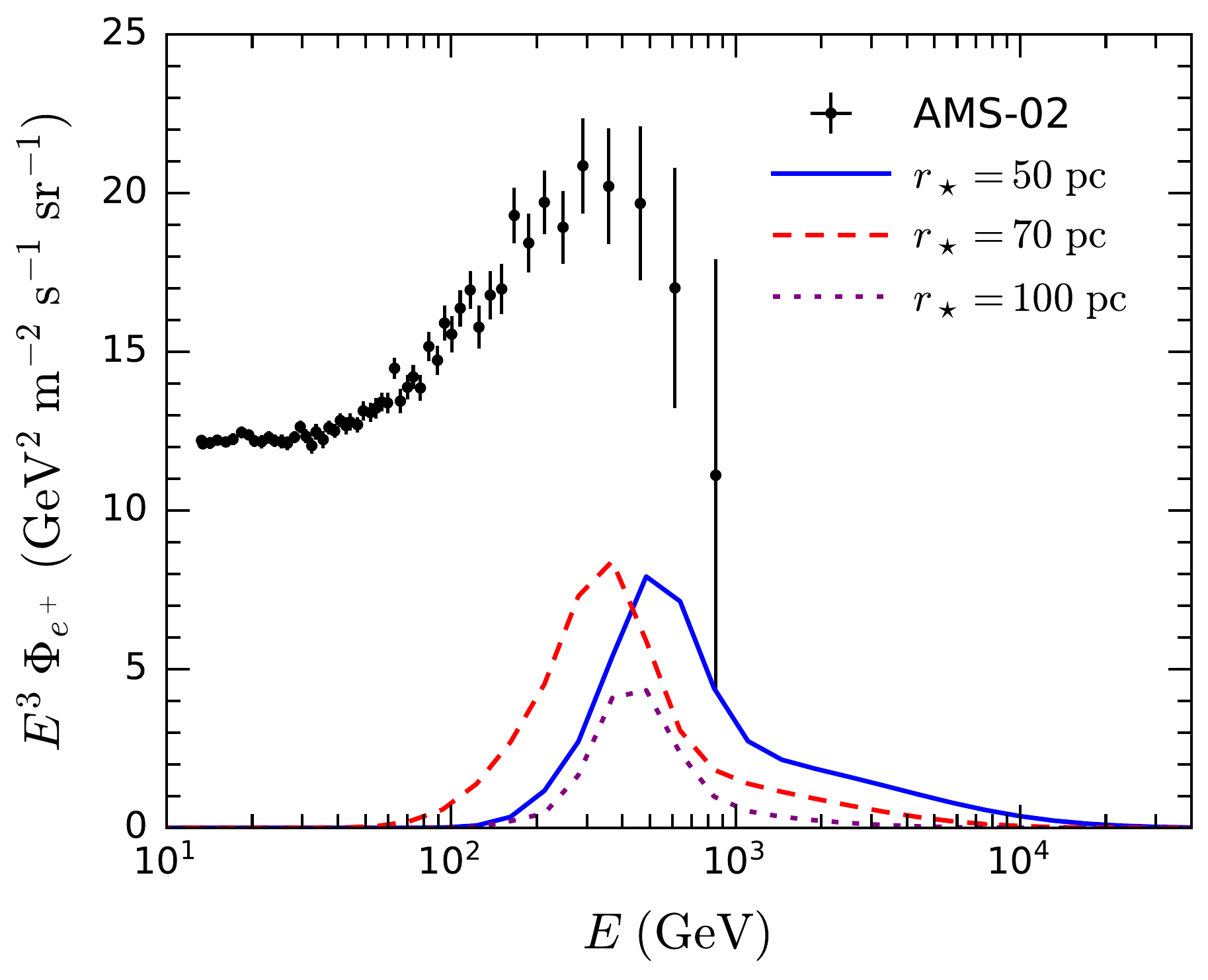}}
\caption{The best results for the $\gamma$-ray spectrum around Geminga (a), the Geminga SBP (b), and the CR positron spectrum (c) assuming $r_\star = 50~\si{pc}$ (blue solid lines), $r_\star = 70~\si{pc}$ (red dashed lines), and $r_\star = 100~\si{pc}$ (purple dotted lines).
In the upper left panel, the upper limits in $5~\si{GeV} \lesssim E_\gamma \lesssim 100~\si{GeV}$ are given by the X19 analysis of Fermi-LAT data~\cite{Xi:2018mii}.
The other experimental data are the same as in Fig.~\ref{fig:DM}.}
\label{fig:Xi}
\end{figure}

In order to get optimistic results, 
we adopt the most loose upper limits on the $\gamma$-ray flux in $10\text{--}500~\si{GeV}$ derived by the X19 analysis, i.e., the upper limits in the upper panel of Fig.~6 in the X19 paper~\cite{Xi:2018mii}.
The parameters of the best results we obtain for $r_\star = 50,~70,~100~\si{pc}$ are listed in Table~\ref{tab:bestpara}.
The corresponding predictions for the $\gamma$-ray spectrum, the SBP, and the positron spectrum are demonstrated in Fig.~\ref{fig:Xi}.
While the HAWC data are properly fitted and the $\gamma$-ray flux in $5~\si{GeV} \lesssim E_\gamma \lesssim 100~\si{GeV}$ lies below the X19 upper limits, we find that Geminga can only supply less than $50\%$ of the AMS-02 positron flux at $E \sim 400~\si{GeV}$.

These results show that the X19 constraint favor $\gamma < 2$, $\eta \lesssim 0.21$, and $E_\mathrm{lc}$ of several hundred GeV, which suppress the $\gamma$-ray flux at $\sim \mathcal{O}(10)~\si{GeV}$.
According to an approximate relation~\cite{Xi:2018mii}
\begin{equation}
E_\gamma = 20 \left(\frac{E}{100~\si{TeV}}\right)^2 \si{TeV}
\end{equation}
for $e^\pm$ ICS off CMB photons, $\mathcal{O}(10)~\si{GeV}$ $\gamma$ rays are induced by $\mathcal{O}(\si{TeV})$ $e^\pm$.
Thus, the X19 constraint implies less $\mathcal{O}(\si{TeV})$ positrons and electrons from Geminga, resulting in a lower CR positron flux for $E \sim \mathcal{O}(100)~\si{GeV}$ at the Earth, which is insufficient to explain the AMS-02 excess.

\section{Summary and discussion}
\label{sec:sum}

In this work, we attempt to explain the AMS-02 positron excess by the nearby pulsar Geminga assuming a two-zone diffusion scenario and an $e^\pm$ injection spectrum with a low energy cutoff, taking into account the $\gamma$-ray data from HAWC and Fermi-LAT.
The analyses of Fermi-LAT data for extended $\gamma$-ray emissions around Geminga by two groups have obtained different results.
While the X19 analysis found no such emission and derive upper limits on the $\gamma$-ray flux, the D19 analysis claimed an observation of the extended $\gamma$ rays.
We have considered both results separately.

By fitting the D19 observation and the HAWC data assuming no low energy cutoff in the injection spectrum, we find that the conversion efficiency $\eta$ should be as large as $60\%$ to account for the AMS-02 positron excess.
Nonetheless, if a low energy cutoff $E_\mathrm{lc} = 20~\si{GeV}$ is introduced, we would only need a $30\%$ conversion efficiency, which is much more realistic.
Therefore, it is plausible to interpret the positron excess by Geminga, according to the D19 analysis.

On the other hand, if the stringent constraint from the X19 analysis is considered, we find that Geminga could not accounting for the total positron excess.
We carry out a scan in the parameter space for the boundary radius $r_\star = 50,~70,~100~\si{pc}$ and require to fit the HAWC data and satisfy the X19 constraint.
The best results we obtain can only explain a faction of the AMS-02 positron flux lower than $50\%$ at $E\sim 400~\si{GeV}$.
This may imply that more nearby pulsars or other sources are needed to interpret the positron excess.

Since the different conclusions obtained above come from the contradictory results of the two Fermi-LAT analyses, it is crucial to know whether result is true.
This may require a more careful data analysis and more Fermi-LAT data.

\begin{acknowledgments}

We thank Kun Fang for providing the code to solve the two-zone propagation equation.
This work is supported in part by the National Natural Science Foundation of China under Grants No.~11875327 and No.~11805288, 
the Fundamental Research Funds for the Central Universities,
and the Sun Yat-Sen University Science Foundation.
Q.Y. is supported by the Program for Innovative Talents and Entrepreneur in Jiangsu.

\end{acknowledgments}

\bibliographystyle{utphys}
\bibliography{bpl}

\providecommand{\href}[2]{#2}\begingroup\raggedright\begin{thebibliography}{10}

\bibitem{PAMELA:2008gwm}
{\bfseries PAMELA} Collaboration, O.~Adriani {\em et~al.}, ``{An anomalous
  positron abundance in cosmic rays with energies 1.5-100 GeV},''
  \href{http://dx.doi.org/10.1038/nature07942}{{\em Nature} {\bfseries 458}
  (2009) 607--609}, \href{http://arxiv.org/abs/0810.4995}{{\ttfamily
  arXiv:0810.4995 [astro-ph]}}.

\bibitem{Fermi-LAT:2011baq}
{\bfseries Fermi-LAT} Collaboration, M.~Ackermann {\em et~al.}, ``{Measurement
  of separate cosmic-ray electron and positron spectra with the Fermi Large
  Area Telescope},''
  \href{http://dx.doi.org/10.1103/PhysRevLett.108.011103}{{\em Phys. Rev.
  Lett.} {\bfseries 108} (2012) 011103},
  \href{http://arxiv.org/abs/1109.0521}{{\ttfamily arXiv:1109.0521
  [astro-ph.HE]}}.

\bibitem{AMS:2013fma}
{\bfseries AMS} Collaboration, M.~Aguilar {\em et~al.}, ``{First Result from
  the Alpha Magnetic Spectrometer on the International Space Station: Precision
  Measurement of the Positron Fraction in Primary Cosmic Rays of
  0.5\textendash{}350 GeV},''
  \href{http://dx.doi.org/10.1103/PhysRevLett.110.141102}{{\em Phys. Rev.
  Lett.} {\bfseries 110} (2013) 141102}.

\bibitem{AMS:2019iwo}
{\bfseries AMS} Collaboration, M.~Aguilar {\em et~al.}, ``{Towards
  Understanding the Origin of Cosmic-Ray Electrons},''
  \href{http://dx.doi.org/10.1103/PhysRevLett.122.101101}{{\em Phys. Rev.
  Lett.} {\bfseries 122} (2019) 101101}.

\bibitem{Bergstrom:2008gr}
L.~Bergstrom, T.~Bringmann, and J.~Edsjo, ``{New Positron Spectral Features
  from Supersymmetric Dark Matter - a Way to Explain the PAMELA Data?},''
  \href{http://dx.doi.org/10.1103/PhysRevD.78.103520}{{\em Phys. Rev. D}
  {\bfseries 78} (2008) 103520},
  \href{http://arxiv.org/abs/0808.3725}{{\ttfamily arXiv:0808.3725
  [astro-ph]}}.

\bibitem{Cholis:2008hb}
I.~Cholis, L.~Goodenough, D.~Hooper, M.~Simet, and N.~Weiner, ``{High Energy
  Positrons From Annihilating Dark Matter},''
  \href{http://dx.doi.org/10.1103/PhysRevD.80.123511}{{\em Phys. Rev. D}
  {\bfseries 80} (2009) 123511},
  \href{http://arxiv.org/abs/0809.1683}{{\ttfamily arXiv:0809.1683 [hep-ph]}}.

\bibitem{Yin:2008bs}
P.-f. Yin, Q.~Yuan, J.~Liu, J.~Zhang, X.-j. Bi, and S.-h. Zhu, ``{PAMELA data
  and leptonically decaying dark matter},''
  \href{http://dx.doi.org/10.1103/PhysRevD.79.023512}{{\em Phys. Rev. D}
  {\bfseries 79} (2009) 023512},
  \href{http://arxiv.org/abs/0811.0176}{{\ttfamily arXiv:0811.0176 [hep-ph]}}.

\bibitem{Yuan:2013eja}
Q.~Yuan, X.-J. Bi, G.-M. Chen, Y.-Q. Guo, S.-J. Lin, and X.~Zhang,
  ``{Implications of the AMS-02 positron fraction in cosmic rays},''
  \href{http://dx.doi.org/10.1016/j.astropartphys.2014.05.005}{{\em Astropart.
  Phys.} {\bfseries 60} (2015) 1--12},
  \href{http://arxiv.org/abs/1304.1482}{{\ttfamily arXiv:1304.1482
  [astro-ph.HE]}}.

\bibitem{Hooper:2008kg}
D.~Hooper, P.~Blasi, and P.~D. Serpico, ``{Pulsars as the Sources of High
  Energy Cosmic Ray Positrons},''
  \href{http://dx.doi.org/10.1088/1475-7516/2009/01/025}{{\em JCAP} {\bfseries
  01} (2009) 025}, \href{http://arxiv.org/abs/0810.1527}{{\ttfamily
  arXiv:0810.1527 [astro-ph]}}.

\bibitem{Yuksel:2008rf}
H.~Yuksel, M.~D. Kistler, and T.~Stanev, ``{TeV Gamma Rays from Geminga and the
  Origin of the GeV Positron Excess},''
  \href{http://dx.doi.org/10.1103/PhysRevLett.103.051101}{{\em Phys. Rev.
  Lett.} {\bfseries 103} (2009) 051101},
  \href{http://arxiv.org/abs/0810.2784}{{\ttfamily arXiv:0810.2784
  [astro-ph]}}.

\bibitem{Yin:2013vaa}
P.-F. Yin, Z.-H. Yu, Q.~Yuan, and X.-J. Bi, ``{Pulsar interpretation for the
  AMS-02 result},'' \href{http://dx.doi.org/10.1103/PhysRevD.88.023001}{{\em
  Phys. Rev. D} {\bfseries 88} (2013) 023001},
  \href{http://arxiv.org/abs/1304.4128}{{\ttfamily arXiv:1304.4128
  [astro-ph.HE]}}.

\bibitem{Feng:2015uta}
J.~Feng and H.-H. Zhang, ``{Pulsar interpretation of lepton spectra measured by
  AMS-02},'' \href{http://dx.doi.org/10.1140/epjc/s10052-016-4092-y}{{\em Eur.
  Phys. J. C} {\bfseries 76} (2016) 229},
  \href{http://arxiv.org/abs/1504.03312}{{\ttfamily arXiv:1504.03312
  [hep-ph]}}.

\bibitem{Hooper:2017tkg}
D.~Hooper and T.~Linden, ``{Measuring the Local Diffusion Coefficient with
  H.E.S.S. Observations of Very High-Energy Electrons},''
  \href{http://dx.doi.org/10.1103/PhysRevD.98.083009}{{\em Phys. Rev. D}
  {\bfseries 98} (2018) 083009},
  \href{http://arxiv.org/abs/1711.07482}{{\ttfamily arXiv:1711.07482
  [astro-ph.HE]}}.

\bibitem{Cholis:2017ccs}
I.~Cholis, T.~Karwal, and M.~Kamionkowski, ``{Features in the Spectrum of
  Cosmic-Ray Positrons from Pulsars},''
  \href{http://dx.doi.org/10.1103/PhysRevD.97.123011}{{\em Phys. Rev. D}
  {\bfseries 97} (2018) 123011},
  \href{http://arxiv.org/abs/1712.00011}{{\ttfamily arXiv:1712.00011
  [astro-ph.HE]}}.

\bibitem{Fang:2018qco}
K.~Fang, X.-J. Bi, P.-F. Yin, and Q.~Yuan, ``{Two-zone diffusion of electrons
  and positrons from Geminga explains the positron anomaly},''
  \href{http://dx.doi.org/10.3847/1538-4357/aad092}{{\em Astrophys. J.}
  {\bfseries 863} (2018) 30}, \href{http://arxiv.org/abs/1803.02640}{{\ttfamily
  arXiv:1803.02640 [astro-ph.HE]}}.

\bibitem{Profumo:2018fmz}
S.~Profumo, J.~Reynoso-Cordova, N.~Kaaz, and M.~Silverman, ``{Lessons from HAWC
  pulsar wind nebulae observations: The diffusion constant is not a constant;
  pulsars remain the likeliest sources of the anomalous positron fraction;
  cosmic rays are trapped for long periods of time in pockets of inefficient
  diffusion},'' \href{http://dx.doi.org/10.1103/PhysRevD.97.123008}{{\em Phys.
  Rev. D} {\bfseries 97} (2018) 123008},
  \href{http://arxiv.org/abs/1803.09731}{{\ttfamily arXiv:1803.09731
  [astro-ph.HE]}}.

\bibitem{Cholis:2018izy}
I.~Cholis, T.~Karwal, and M.~Kamionkowski, ``{Studying the Milky Way pulsar
  population with cosmic-ray leptons},''
  \href{http://dx.doi.org/10.1103/PhysRevD.98.063008}{{\em Phys. Rev. D}
  {\bfseries 98} (2018) 063008},
  \href{http://arxiv.org/abs/1807.05230}{{\ttfamily arXiv:1807.05230
  [astro-ph.HE]}}.

\bibitem{Tang:2018wyr}
X.~Tang and T.~Piran, ``{Positron flux and $\gamma$-ray emission from Geminga
  pulsar and pulsar wind nebula},''
  \href{http://dx.doi.org/10.1093/mnras/stz268}{{\em Mon. Not. Roy. Astron.
  Soc.} {\bfseries 484} (2019) 3491--3501},
  \href{http://arxiv.org/abs/1808.02445}{{\ttfamily arXiv:1808.02445
  [astro-ph.HE]}}.

\bibitem{Xi:2018mii}
S.-Q. Xi, R.-Y. Liu, Z.-Q. Huang, K.~Fang, and X.-Y. Wang, ``{GeV observations
  of the extended pulsar wind nebulae constrain the pulsar interpretations of
  the cosmic-ray positron excess},''
  \href{http://dx.doi.org/10.3847/1538-4357/ab20c9}{{\em Astrophys. J.}
  {\bfseries 878} (2019) 104},
  \href{http://arxiv.org/abs/1810.10928}{{\ttfamily arXiv:1810.10928
  [astro-ph.HE]}}.

\bibitem{Johannesson:2019jlk}
G.~Johannesson, T.~A. Porter, and I.~V. Moskalenko, ``{Cosmic-Ray Propagation
  in Light of the Recent Observation of Geminga},''
  \href{http://dx.doi.org/10.3847/1538-4357/ab258e}{{\em Astrophys. J.}
  {\bfseries 879} (2019) 91}, \href{http://arxiv.org/abs/1903.05509}{{\ttfamily
  arXiv:1903.05509 [astro-ph.HE]}}.

\bibitem{DiMauro:2019yvh}
M.~Di~Mauro, S.~Manconi, and F.~Donato, ``{Detection of a $\gamma$-ray halo
  around Geminga with the Fermi -LAT data and implications for the positron
  flux},'' \href{http://dx.doi.org/10.1103/PhysRevD.104.089903}{{\em Phys. Rev.
  D} {\bfseries 100} (2019) 123015},
  \href{http://arxiv.org/abs/1903.05647}{{\ttfamily arXiv:1903.05647
  [astro-ph.HE]}}. [Erratum: Phys.Rev.D 104, 089903 (2021)].

\bibitem{Fang:2019ayz}
K.~Fang, X.-J. Bi, and P.-F. Yin, ``{Reanalyze the pulsar scenario to explain
  the cosmic positron excess considering the recent developments},''
  \href{http://dx.doi.org/10.3847/1538-4357/ab3fac}{{\em Astrophys. J.}
  {\bfseries 884} (2019) 124--128},
  \href{http://arxiv.org/abs/1906.08542}{{\ttfamily arXiv:1906.08542
  [astro-ph.HE]}}.

\bibitem{Manconi:2020ipm}
S.~Manconi, M.~Di~Mauro, and F.~Donato, ``{Contribution of pulsars to
  cosmic-ray positrons in light of recent observation of inverse-Compton
  halos},'' \href{http://dx.doi.org/10.1103/PhysRevD.102.023015}{{\em Phys.
  Rev. D} {\bfseries 102} (2020) 023015},
  \href{http://arxiv.org/abs/2001.09985}{{\ttfamily arXiv:2001.09985
  [astro-ph.HE]}}.

\bibitem{Wang:2021xph}
S.-H. Wang, K.~Fang, X.-J. Bi, and P.-F. Yin, ``{Test of the superdiffusion
  model in the interstellar medium around the Geminga pulsar},''
  \href{http://dx.doi.org/10.1103/PhysRevD.103.063035}{{\em Phys. Rev. D}
  {\bfseries 103} (2021) 063035},
  \href{http://arxiv.org/abs/2101.01438}{{\ttfamily arXiv:2101.01438
  [astro-ph.HE]}}.

\bibitem{Fang:2022mdg}
K.~Fang and X.-J. Bi, ``{Interpretation of the puzzling gamma-ray spectrum of
  the Geminga halo},''
  \href{http://dx.doi.org/10.1103/PhysRevD.105.103007}{{\em Phys. Rev. D}
  {\bfseries 105} (2022) 103007},
  \href{http://arxiv.org/abs/2203.01546}{{\ttfamily arXiv:2203.01546
  [astro-ph.HE]}}.

\bibitem{HAWC:2017kbo}
{\bfseries HAWC} Collaboration, A.~U. Abeysekara {\em et~al.}, ``{Extended
  gamma-ray sources around pulsars constrain the origin of the positron flux at
  Earth},'' \href{http://dx.doi.org/10.1126/science.aan4880}{{\em Science}
  {\bfseries 358} (2017) 911--914},
  \href{http://arxiv.org/abs/1711.06223}{{\ttfamily arXiv:1711.06223
  [astro-ph.HE]}}.

\bibitem{LHAASO:2021crt}
{\bfseries LHAASO} Collaboration, F.~Aharonian {\em et~al.}, ``{Extended
  Very-High-Energy Gamma-Ray Emission Surrounding PSR J0622+3749 Observed by
  LHAASO-KM2A},'' \href{http://dx.doi.org/10.1103/PhysRevLett.126.241103}{{\em
  Phys. Rev. Lett.} {\bfseries 126} (2021) 241103},
  \href{http://arxiv.org/abs/2106.09396}{{\ttfamily arXiv:2106.09396
  [astro-ph.HE]}}.

\bibitem{Bao:2021hey}
L.-Z. Bao, K.~Fang, and X.-J. Bi, ``{Slow diffusion is necessary to explain the
  gamma-ray pulsar halos},'' \href{http://arxiv.org/abs/2107.07395}{{\ttfamily
  arXiv:2107.07395 [astro-ph.HE]}}.

\bibitem{Fichtel:1975}
C.~E. {Fichtel}, R.~C. {Hartman}, D.~A. {Kniffen}, D.~J. {Thompson}, G.~F.
  {Bignami}, H.~{{\"O}gelman}, M.~E. {{\"O}zel}, and T.~{T{\"u}mer},
  ``{High-energy gamma-ray results from the second Small Astronomy
  Satellite.},'' \href{http://dx.doi.org/10.1086/153590}{{\em \apj} {\bfseries
  198} (1975) 163--182}.

\bibitem{Manchester:2004bp}
R.~N. Manchester, G.~B. Hobbs, A.~Teoh, and M.~Hobbs, ``{The Australia
  Telescope National Facility pulsar catalogue},''
  \href{http://dx.doi.org/10.1086/428488}{{\em Astron. J.} {\bfseries 129}
  (2005) 1993}, \href{http://arxiv.org/abs/astro-ph/0412641}{{\ttfamily
  arXiv:astro-ph/0412641}}.

\bibitem{Faherty:2007}
J.~Faherty, F.~Walter, and J.~Anderson, ``{The trigonometric parallax of the
  neutron star Geminga},''
  \href{http://dx.doi.org/10.1007/s10509-007-9368-0}{{\em Astrophys. Space
  Sci.} {\bfseries 308} (2007) 225–230}.

\bibitem{Crusius:1988}
A.~{Crusius} and R.~{Schlickeiser}, ``{Synchrotron radiation in a thermal
  plasma with large-scale random magnetic fields},'' {\em Astron. Astrophys.}
  {\bfseries 196} (1988) 327--337.

\bibitem{Fang:2020dmi}
K.~Fang, X.-J. Bi, S.-J. Lin, and Q.~Yuan, ``{Klein\textendash{}Nishina Effect
  and the Cosmic Ray Electron Spectrum},''
  \href{http://dx.doi.org/10.1088/0256-307X/38/3/039801}{{\em Chin. Phys.
  Lett.} {\bfseries 38} (2021) 039801},
  \href{http://arxiv.org/abs/2007.15601}{{\ttfamily arXiv:2007.15601
  [astro-ph.HE]}}.

\bibitem{Kolmogorov:1941}
A.~N. Kolmogorov, ``{The local structure of turbulence in incompressible
  viscous fluid for very large Reynolds numbers},'' {\em C. R. Acad. Sci. URSS}
  {\bfseries 30} (1941) 301--305.

\bibitem{Moskalenko:1997gh}
I.~V. Moskalenko and A.~W. Strong, ``{Production and propagation of cosmic ray
  positrons and electrons},'' \href{http://dx.doi.org/10.1086/305152}{{\em
  Astrophys. J.} {\bfseries 493} (1998) 694--707},
  \href{http://arxiv.org/abs/astro-ph/9710124}{{\ttfamily
  arXiv:astro-ph/9710124}}.

\bibitem{Fang:2007sc}
J.~Fang and L.~Zhang, ``{Non-thermal emission from old supernova remnants},''
  \href{http://dx.doi.org/10.1111/j.1365-2966.2007.12766.x}{{\em Mon. Not. Roy.
  Astron. Soc.} {\bfseries 384} (2008) 1119},
  \href{http://arxiv.org/abs/0711.4173}{{\ttfamily arXiv:0711.4173
  [astro-ph]}}.

\bibitem{Liu:2019sfl}
R.-Y. Liu, C.~Ge, X.-N. Sun, and X.-Y. Wang, ``{Constraining the Magnetic Field
  in the TeV Halo of Geminga with X-Ray Observations},''
  \href{http://dx.doi.org/10.3847/1538-4357/ab125c}{{\em Astrophys. J.}
  {\bfseries 875} (2019) 149},
  \href{http://arxiv.org/abs/1904.11438}{{\ttfamily arXiv:1904.11438
  [astro-ph.HE]}}.

\bibitem{Feroz:2008xx}
F.~Feroz, M.~P. Hobson, and M.~Bridges, ``{MultiNest: an efficient and robust
  Bayesian inference tool for cosmology and particle physics},''
  \href{http://dx.doi.org/10.1111/j.1365-2966.2009.14548.x}{{\em Mon. Not. Roy.
  Astron. Soc.} {\bfseries 398} (2009) 1601--1614},
  \href{http://arxiv.org/abs/0809.3437}{{\ttfamily arXiv:0809.3437
  [astro-ph]}}.

\end{thebibliography}\endgroup

\end{document}